\documentclass[manuscript,screen]{acmart}
\usepackage{graphicx}
\usepackage{amsmath}
\usepackage{pifont}
\usepackage{booktabs}
\usepackage{microtype}
\usepackage{multicol}
\usepackage{multirow}
\usepackage{xspace}
\usepackage{xcolor}
\newcommand{\cmark}{\color{green}\ding{51}}
\newcommand{\xmark}{\color{red}\ding{55}}
\newcommand{\appName}{\textit{Discussion Jockey 2}\xspace}

\AtBeginDocument{%
  }

\setcopyright{acmlicensed}
\copyrightyear{2024}
\acmYear{2024}
\acmDOI{XXXXXXX.XXXXXXX}

\acmConference[Conference acronym '24]{Make sure to enter the correct
  conference title from your rights confirmation email}{June 03--05,
  2024}{Woodstock, NY}

\acmISBN{978-1-4503-XXXX-X/2024/06}

\usepackage{amssymb}

\begin{document}
\title{Augmenting Online Meetings with Context-Aware Real-time Music Generation}

\author{Haruki Suzawa}
\orcid{0000-0002-3067-8897}
\email{harukisuzawa@gmail.com}
\affiliation{%
  \institution{Nara Institute of Science and Technology}
  \city{Nara}
  \country{Japan}
}

\author{Ko Watanabe}
\orcid{0000-0003-0252-1785}
\email{ko.watanabe@dfki.de}
\affiliation{%
  \institution{German Research Center for Artificial Intelligence (DFKI)}
  \city{Kaiserslautern}
  \country{Germany}
}

\author{Andreas Dengel}
\orcid{0000-0002-6100-8255}
\email{andreas.dengel@dfki.de}
\affiliation{%
  \institution{German Research Center for Artificial Intelligence (DFKI)}
  \city{Kaiserslautern}
  \country{Germany}
}

\author{Shoya Ishimaru}
\orcid{0000-0002-5374-1510}
\email{ishimaru@omu.ac.jp}
\affiliation{%
  \institution{Osaka Metropolitan University}
  \city{Osaka}
  \country{Japan}
}

\renewcommand{\shortauthors}{Suzawa et al.}

\begin{abstract}
As online communication continues to expand, participants often face cognitive fatigue and reduced engagement. Cognitive augmentation, which leverages technology to enhance human abilities, offers promising solutions to these challenges. In this study, we investigate the potential of generative artificial intelligence (GenAI) for real-time music generation to enrich online meetings. We introduce \textit{Discussion Jockey 2}, a system that dynamically produces background music in response to live conversation transcripts. Through a user study involving 14 participants in an online interview setting, we examine the system's impact on relaxation, concentration, and overall user experience. The findings reveal that AI-generated background music significantly enhances user relaxation (average score: 5.75/9) and concentration (average score: 5.86/9). This research underscores the promise of context-aware music generation in improving the quality of online communication and points to future directions for optimizing its implementation across various virtual environments.
\end{abstract}




\received{20 February 2024}
\received[revised]{12 March 2024}
\received[accepted]{5 June 2024}

\maketitle

\section{Introduction}
Cognitive augmentation leverages technology to enhance human cognitive abilities~\cite{schmidt2017augmenting, clinch2023augmented}. 
This concept is particularly applied in mobile computing devices~\cite{schmidt2017augmenting, hypermind2018ishimaru, yamaoka2022experience, yamaoka2025img2vocab}.
Our society now operates in a hybrid mode, where offline and online communication blend physical and virtual interactions. 
This transition is driven by technological advancements and the growing need for flexible ways to connect and collaborate. 
In this new era, cognitive augmentation holds significant potential for enhancing communication.
For example, in online communication, cognitive augmentation tools are used to support memory retention~\cite{mizuho2023virtual}, boost engagement~\cite{watanabe2023engauge, watanabe2024metacognition}, facilitate smooth interactions~\cite{suzawa2022supporting,kyto2021from}, and assist in decision-making processes~\cite{kim2024engaged}. 
These tools range from simple note-taking applications~\cite{son2023okay} that enhance memory retention to advanced AI-driven systems providing real-time feedback and suggestions~\cite{rajaram2024blenderscape}, transforming how we conduct and participate in meetings.

Recent generative artificial intelligence (GenAI) advancements offer new possibilities for augmenting meetings~\cite{rajaram2024blenderscape, park2024coexplorer}. 
For example, \citet{son2023okay} investigated real-time transcript summarization during meetings using large language models (LLMs) like BERT~\cite{kenton2019bert}. 
The goal was to see if such summarization could help participants focus on the meeting content. 
The results show that real-time transcription summarization with LLMs helps participants maintain focus. 
Another GenAI application is presented by \citet{rajaram2024blenderscape}, who developed a system called \textit{BlendScape} that uses GenAI to personalize video-conferencing environments. 
This application collects transcripts during meetings and uses them as prompts to generate background images, creating a customized video-conferencing environment for an immersive experience. 
Among these approaches, real-time music generation remains unexplored in the context of augmenting meetings.

Music profoundly impacts human emotions and can be a powerful tool for healing and relaxation~\cite{conrad2010music, feng2024co}.
The right music can create a calming atmosphere, reduce anxiety, and improve overall mood~\cite{mccraty1998effects}. Background music can also help improve focus and concentration~\cite{woods2024rapid}.
Concerning the previous findings, our study aims to use music GenAI to consider the context for creating a personalized and practical auditory experience—the strategic use of music in online interviews and discovering how beneficial for humans.

In this study, we focus on evaluating the effectiveness of GenAI for music generation in online meetings.
Our target scenario is an online interview setting where the interviewer and interviewee are in different locations.
By collecting transcripts during the online interview, we generate music that considers the context of the interview and personalizes the music.
Our demonstration aims to explore the following research questions: 
(RQ1) Does context-aware music generation make participants feel relaxed in online meetings?
(RQ2) Does context-aware music generation enhance participants' concentration in online meetings?

\begin{table}[t!]
  \centering
  \renewcommand{\arraystretch}{1.0}
  \caption{Comparison of related work for augmenting online meetings. The ``Context Aware'' column represents a checklist indicating whether the research utilizes the meeting transcript. The ``GenAI'' column represents a checklist indicating whether the research use GenAI technology. The ``Music'' column represents a checklist indicating whether the research focuses on using music in meetings.}
  \scalebox{0.84}{
    \begin{tabular}{lcccl}
      \hline
        \textbf{Author}                 & \textbf{Context Aware} & \textbf{GenAI} & \textbf{Music} & \textbf{Performance Detail} \\ 
        \hline
        \citet{son2023okay}             & \cmark                 & \cmark & \xmark            & Real-time summarization of the meeting transcripts using LLMs (BERT). \\ 
        \citet{park2024coexplorer}      & \cmark                 & \cmark & \xmark            & Introduce \textit{CoExplorer2D} and \textit{CoExplorerVR} managing meeting progress. \\
        \citet{han2024when}             & \cmark                 & \cmark & \xmark            & Collaborative creation of concept-based image generation (Midjourney). \\ 
        \citet{rajaram2024blenderscape} & \cmark                 & \cmark & \xmark            & Real-time generation of a meeting background images. \\ 
        \citet{feng2024co}              & \xmark                 & \xmark & \cmark             & Collaborative music creation for therapy using the prototype \textit{ComString}. \\ 
        \citet{suzawa2022supporting}    & \xmark                 & \xmark & \cmark             & A different metronome (BPM) is sounded in real-time for each participant. \\ 
        \textbf{Ours}                   & \cmark                 & \cmark & \cmark             & Real-time generation of a personalized music using transcript as a prompt.    \\ 
        \hline
      \end{tabular}
  }
  \label{tb:research-position}
\end{table}

\section{Related Work}
Table~\ref{tb:research-position} highlights the uniqueness of our research in comparison to existing studies. Several researchers have explored context-aware meeting augmentation~\cite{son2023okay, park2024coexplorer, han2024when, rajaram2024blenderscape}. For instance, \citet{son2023okay} and \citet{park2024coexplorer} concentrate on summarizing or facilitating meetings through transcripts, utilizing LLMs to generate summaries or assist in meeting progression. Conversely, \citet{han2024when} and \citet{rajaram2024blenderscape} employ image generation to enhance meetings. Specifically, \citet{han2024when} use GenAI for real-time collaborative image creation during discussions, while \citet{rajaram2024blenderscape} generate background images based on speech context to enhance immersion. Although these studies utilize GenAI, none focus on context-aware music generation for online meeting augmentation.

In the realm of music used during online meetings, \citet{feng2024co} propose the collaborative creation of therapeutic music using \textit{ComString}, which involves Collaborative Digital Musical Instruments (CDMIs) for real-time music composition. Their work aims to generate music for therapy, not general meeting contexts, and does not incorporate transcripts. Our research closely relates to \citet{suzawa2022supporting}, where metronome (BPM) music is produced based on participants' speech data, with higher BPM for more active speakers and lower for quieter ones. This study aims to balance speaking time among participants but does not focus on the context or content of the transcripts. Both research uses music as a target in the meeting, but neither work uses transcripts nor GenAI.

In conclusion, while various researchers have explored context-aware online meeting augmentation and the use of music in meetings, the context-aware generation of music for meeting enhancement remains underexplored. Therefore, our research aims to investigate the impact of applying context-aware music generation in online meetings.



\begin{figure}[t!]
  \centering
  \begin{minipage}{0.54\linewidth}
    \centering
    \includegraphics[width=\linewidth]{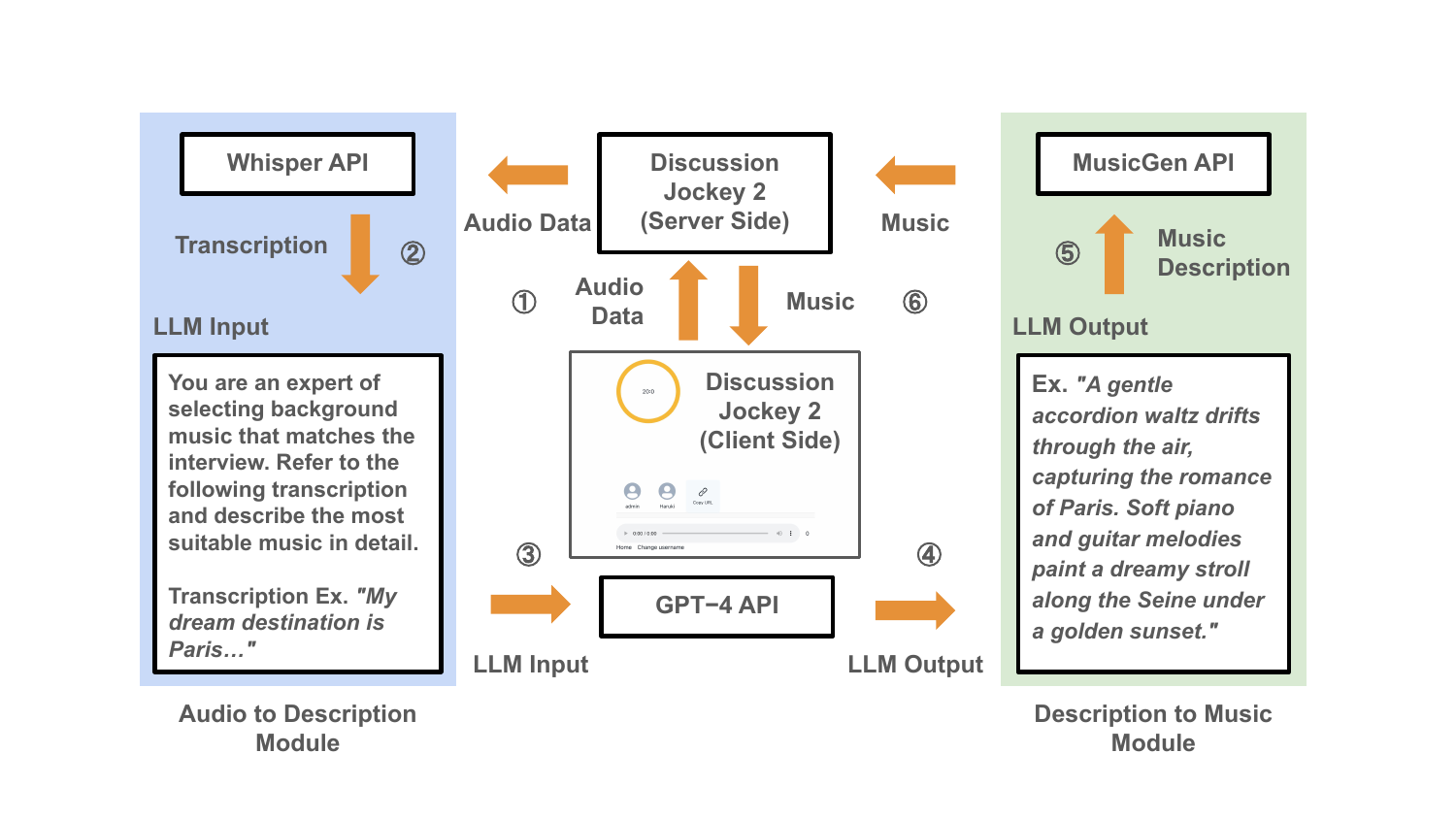}
    \caption{Proposed architecture of \appName. The application uses Whisper API to collect speech transcripts. The transcript and template prompt are input in GPT-4 to generate a music description prompt optimized for MusicGen API. The generated music is then feedback to the application and played in the participant interface.}
    \label{fig:dj2_system}
  \end{minipage}
  \hfill
  \begin{minipage}{0.422\linewidth}
    \centering
    \includegraphics[width=\linewidth]{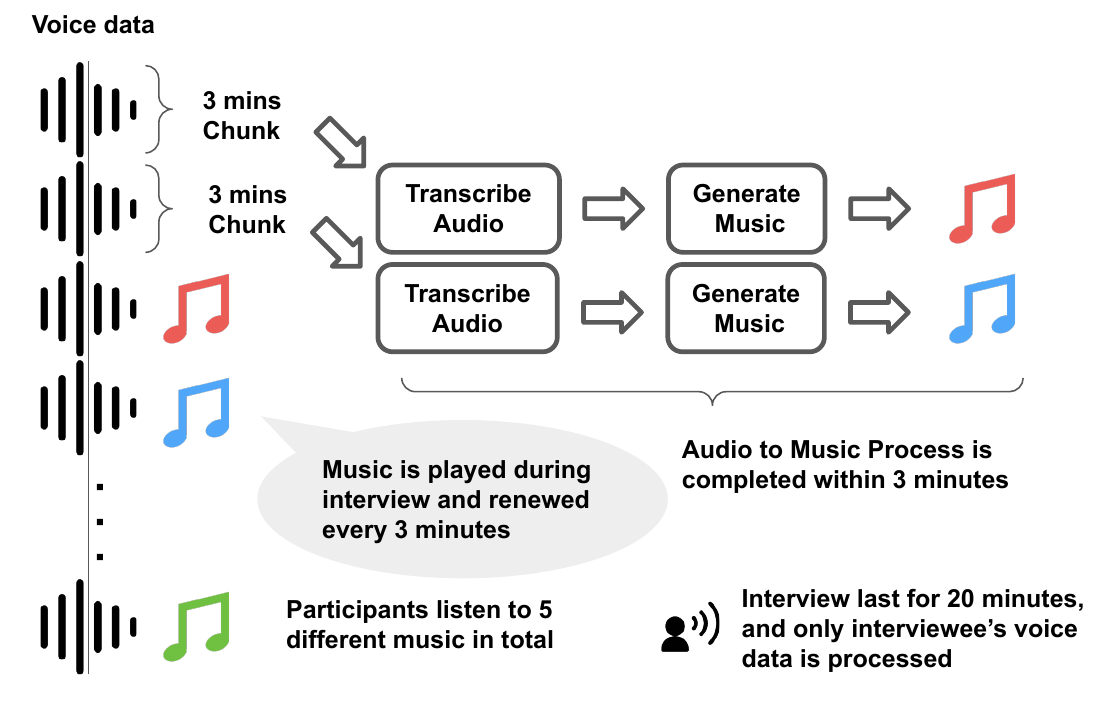}
    \caption{Experiment Design: The music is generated using transcription from the first three minutes of the interview. While the music is played, the new three-minute transcription will used for generation. The procedure will be repeated until five different pieces of music are played.}
    \label{fig:experiment}
  \end{minipage}
\end{figure}

\section{Methodology}
Figure~\ref{fig:dj2_system} illustrates the architecture of \appName, a system that generates real-time background music based on live conversation. The client-side, built with React and WebSocket, facilitates seamless user-server interaction. The workflow of the server-side consists of three stages: (1) transcribing real-time audio using Whisper API~\footnote{\url{https://openai.com/research/whisper}}, (2) generating a music description via GPT-4 after accumulating three minutes of transcription, and (3) creating a customized music track with MusicGen API~\footnote{\url{https://replicate.com/meta/musicgen/api}}. The description defines tempo, style, and instrumentation, ensuring dynamic adaptation to conversation context. To reduce music generation time, the music length was set to 10 seconds and looped for 3 minutes. For instance, when the transcription includes \textit{``My dream destination is Paris...''}, the system generates a fitting music description, as shown in Figure~\ref{fig:dj2_system}: \textit{``A gentle accordion waltz drifts through the air, capturing the romance of Paris. Soft piano and guitar melodies paint a dreamy stroll along the Seine under a golden sunset.''}.

We conducted a 20-minute one-on-one interview experiment with 14 participants, as shown in Figure~\ref{fig:experiment}. Only the interviewee's voice data was processed. Background music was introduced six minutes in and changed every three minutes, allowing participants to experience five different pieces. To maintain a natural and engaging conversation, they were asked casual questions about hobbies, travel, and favorite foods. After listening to all five pieces, they evaluated each based on three criteria: whether it helped them relax, improved concentration, and whether they liked it.

\section{Result \& Discussion}
To explore the research questions (RQs) and identify the most beneficial scenarios for the proposed system, we analyzed subjective evaluations of the generated music. 
The average ratings were 5.75 for relaxation (1: very nervous, 9: very relaxed) and 5.86 for concentration (1: very distracted, 9: very concentrated), indicating a tendency to enhance both.
In this experiment, the system generated music that matched the conversation content (Figure~\ref{fig:dj2_system}, LLM Input). 
However, modifying the prompt can generate music tailored to enhance relaxation or concentration.


The average response to ``Did you like the music?'' (1: hated, 9: liked) was 5.67, indicating a generally positive reception. 
A follow-up survey collected 15 responses, with eight citing musical preferences (tune, tempo, melody, mood) as key factors. 
Participants also wanted their preferences for personalized music to be considered in advance. 
The next prototype will incorporate genre, nationality, gender, and age to assess the impact on concentration and relaxation during meetings. 
Three participants prioritized relaxation, while two emphasized the importance of situational and emotional alignment. 
Currently, the system generates music based on transcriptions from two chunks earlier (each three minutes long) to allow for processing time. 
However, enhancing real-time performance will be a critical area for future improvement.
Additionally, two participants prioritized concentration, underscoring the need for non-disruptive music. 
As for volume, responses to ``How loud was the sound played?'' (1: too quiet, 10: too loud) averaged 6.60, indicating that the volume was relatively high and may have influenced responses. 
Meanwhile, ``Did you find the frequency of music transitions comfortable?'' (1: very uncomfortable, 10: very comfortable) received an average score of 6.65, suggesting that the three-minute update interval was generally well-received.

To understand suitable usage scenarios for this system, participants were asked: ``In what situations would you like to use this system?''. A total of 13 responses were collected. The most common category (six responses) involved individual tasks like working, studying, driving, and meditating, with an emphasis on using the system for boring tasks. Reading was specifically mentioned, suggesting future research on generating music based on book content. 
Four responses favored using the system in meetings, with two highlighting its role in setting the mood for casual conversations with friends. Two respondents offered potential educational uses: one suggested using it for relaxation during presentation, and the other advocated for maintaining focus during lectures—both underscoring its value in an educational context.



\section{Conclusion}
This study introduced \textit{Discussion Jockey 2}, a novel approach to augmenting online meetings through real-time, context-aware music generation. By leveraging GenAI technologies, our system personalizes background music based on meeting transcripts, aiming to improve relaxation and concentration during virtual interactions. Our findings suggest that AI-generated music can create a more engaging and comfortable environment, with most participants reporting positive effects. However, personalization and real-time processing remain key factors in optimizing user experience, as individual preferences significantly influence perception. Future work will focus on refining music customization based on user-specific attributes, such as nationality, age, and music preferences, to enhance the adaptability and impact of AI-driven auditory augmentation in virtual meetings.

\begin{acks}
  This work is supported by the OMU Project ``Verifying the Health-Promoting Effects of Music as a Social Prescription''.
\end{acks}

\bibliographystyle{plain}
\bibliography{main}

\end{document}